\documentclass{aa}  

\usepackage{natbib}
\bibpunct{(}{)}{;}{a}{}{,} 

 \usepackage[varg]{txfonts}
\usepackage{graphicx}
 \usepackage{lscape}
 
\usepackage{txfonts}
\usepackage{float} 

\begin{document} 

   \title{Center-to-limb polarization in continuum spectra of F, G, K stars}

   \author{N.M. Kostogryz
          \inst{1,2}
          \and
          S.V. Berdyugina\inst{1, 3}
        }

   \institute{Kiepenheuer-Institut f\"ur Sonnenphysik (KIS), 
              Sch\"oneckstrasse 6, D-79104 Freiburg\\
              \email{kostogryz@kis.uni-freiburg.de, sveta@kis.uni-freiburg.de}
         \and
             Main Astronomical Observatory of NAS of Ukraine,
	     Zabolotnoho str. 27, 03680, Kyiv
	 \and
	 NASA Astrobiology Institute, Institute for Astronomy, University of Hawaii, USA\\
             }

   \date{Received August, 2014; accepted }

 
  \abstract
   { Scattering and absorption processes in stellar atmosphere affect the center-to-limb variations of the intensity (CLVI) 
   and the linear polarization (CLVP) of stellar radiation. 
    }
   {There are several theoretical and observational studies of CLVI using different stellar models, 
    however, most studies of CLVP have concentrated on the solar atmosphere and have not considered the CLVP
    in cooler non-gray stellar atmospheres at all. In this paper, we present
   a theoretical study of the CLV of the intensity and the linear polarization in continuum spectra 
   of different spectral type stars. }
   {We solve the radiative transfer equations for polarized light iteratively assuming no magnetic field 
   and considering a plane-parallel model atmospheres and various opacities.}
   {We calculate the CLVI and the CLVP for Phoenix stellar model atmospheres for 
   the range of effective temperatures (4500 K - 6900 K), gravities ($\log g = 3.0 - 5.0$), and wavelengths 
   (4000 - 7000 $\AA$), which are tabulated and available at the Strasbourg astronomical Data Center (CDS). 
  In addition, we present several tests of our code and compare our results with measurements and calculations
  of CLVI and the CLVP for the Sun.
  The resulting CLVI are fitted with polynomials and their 
  coefficients are presented in this paper.}
   {For the stellar model atmospheres with lower gravity and effective temperature 
   the CLVP is larger.}


   \maketitle
%

\section{Introduction}

Stellar intrinsic polarization from scattering is an important effect for investigating the 
physical and geometrical properties of stars and stellar environments.
The first theoretical prediction of linear polarization of  continuous light in the 
emergent radiation from the early-type stars was made by \cite{chandr46}. 
He solved the radiative transfer equation for a purely scattering atmosphere in radiative 
equilibrium and showed that polarization from the light scattered at the limb of the stellar 
disk is considerable, $\sim12\%$, and could be detected under favourable conditions.
Later, \citet{rucinski70} showed that the high value of limb polarization predicted by Chandrasekhar 
could be only considered as the upper limit  for early-type stars, while polarization for cooler stars would be much smaller. 
\citet{harcol68} demonstrated that rotation distortion (early-type stars are usually fast rotators) and 
limb darkening affect the symmetry of early-type stars and 
should produce detectable polarization under suitable conditions and geometry. 
However, Thomson scattering
is the main source of scattering opacity in the early spectral type stars, but not for the solar or the late-type stars \citep[see,][]{clarke10}.  

The best representative candidate of the solar-type stars is our Sun. As the polarized spectrum formed 
by coherent scattering is a rich source of information about the atmosphere, the scientific 
community has paid a lot of attention to studying the polarized spectrum of the Sun \citep[e.g.][and many others]{fluri99,
 berdyugina02, stenflo05, trujillo09}.
This spectrum was called  the "Second Solar Spectrum" \citep{ivanov91, stenflokeller97}.
It is characterized by a polarized continuous background on which a rich variety  of both 
intrinsically polarized and depolarized lines are superimposed.  Polarized lines look like 
"emission" and depolarized lines look similar to "absorption" lines with respect to continuum polarization.
However, many
lines are also weakly polarized or depolarized by magnetic fields due to the Hanle effect. 
The polarization in the lines and the continuum are  usually on the same order of magnitude, 
and a common zero level should be used as the reference for the line polarization.
There are several observational studies of center-to-limb variation of linear polarization
in different spectral lines and adjacent continua for the Sun \citep{leroy72, mickeyorrall74, wiehr75, 
wiehrbiande03, stenflo05}.
The maximum 
continuum polarization on the Sun is up to $1.7\%$ at the limb for wavelength $3000\AA$ \citep{stenflo05}.
\citet{wiehrbiande03} measured the CLVP in continuum of $0.12\%$ in two wavelength intervals ($3 \AA$) overlapping
each other near $4506\AA$ where no significant line polarization occurs. These measurements are
very difficult since the continuum limb polarization is superposed with strong intensity 
gradient at the limb, and the absolute value of polarization is hard to measure.  
Knowing the degree 
of polarization in the continuum from theoretical predictions helps to interpret the observations. 

As the solar disk can be resolved, it is possible to detect the limb polarization \citep{gand00, gand02, gand05} 
and the center-to-limb variation of polarized light \citep{leroy72, mickeyorrall74, wiehr75, wiehrbiande03} directly. 
In most cases, however, it is not possible to resolve a stellar disk to measure the CLVP of its radiation. 
It is expected that any intrinsic polarization of 
solar-type stars integrated over the disk is likely to be very small, and it can only be increased 
if the symmetry of the disk is broken. This is the case if a star possesses a nonspherical radiation 
field as a result of geometric distortion, for example, due to fast rotation or tides in binaries, or if it has a nonuniform photospheric surface 
brightness. For example, the latter can happen in the case of starspots or inhomogeneities in the outer atmosphere 
where scattering takes place, or because of a transiting exoplanet that blocks a part of the stellar radiation
and, hence, breaks the symmetry of the stellar disk.   

\citet{carciofi05}, \citet{kostogryz11a, kostogryz11b}, and \citet{frantseva12}
attempted to estimate the polarization signal of a star due to the symmetry breaking 
effect that appears when a planet transits the stellar disk. They showed that the polarization degree 
of the transiting exoplanetary system is very sensitive to the intrinsic center-to-limb polarization 
of the host star, however, they used simplified approximations for the stellar limb polarization. In this paper, 
we present a more realistic  theoretical calculation of polarized light 
in stellar atmospheres by solving the radiative transfer equation for the visible spectra of F, G, and K stars.
In section 2 we focus on the theoretical approach to our calculations and on the opacities in the
stellar model atmosphere. Section 3 presents the results of our calculations of CLVI and CLVP for 
different stellar model atmospheres. In addition, we present several tests of our
computer code. Finally, we summarize our findings in section 4.
  

\section{Theoretical approach to continuum polarization calculations}

\subsection {Stellar models and opacities.} 
Models of stellar atmospheres are the most important input in our calculations. 
In our simulations we used Phoenix local thermodynamic equilibrium (LTE) models \citep{hauschildt99} for the range of the effective 
temperatures from 4500 K to 6900 K and the surface gravity $\log g$ from 3.0 to 5.0. 
In addition to 
the Phoenix models, we employed several semiempirical solar model atmospheres 
of the quiet Sun, such as FALC (averaged quiet Sun), FALA (the supergranular cell center), 
FALP (the plage model) 
\citep{fontenla93} and HSRA (averaged quiet Sun) \citep{gingerich71}. 

We calculated the opacities for different wavelengths using the code 
 SLOC \citep{berdyugina91} for stars of different spectral types, assuming solar abundances and metallicity. 
 For the calculation of the stellar continuum opacities we have taken the following contributors into account:
 \begin{itemize}
\item Scattering opacity: 
Thomson scattering on free electrons $e^-$ and Rayleigh scattering on $\rm{H I}$, $\rm{He I}$, $\rm{H_2}$, $\rm{CO}$, $\rm{H_2O}$, and other molecules.\\
\item  Absorption opacity: free-free (ff) and bound-free (bf) transitions
 in $\rm{H^-}$, $\rm{H I}$, $\rm{He I}$, $\rm{He^-}$, $\rm{H_2^-}$, $\rm{H_2^+}$, and metal photoionization. 
\end{itemize}

  \begin{figure*}
  \centering
\begin{minipage}{1.0\textwidth}
  \centering
  \includegraphics[width=.45\linewidth]{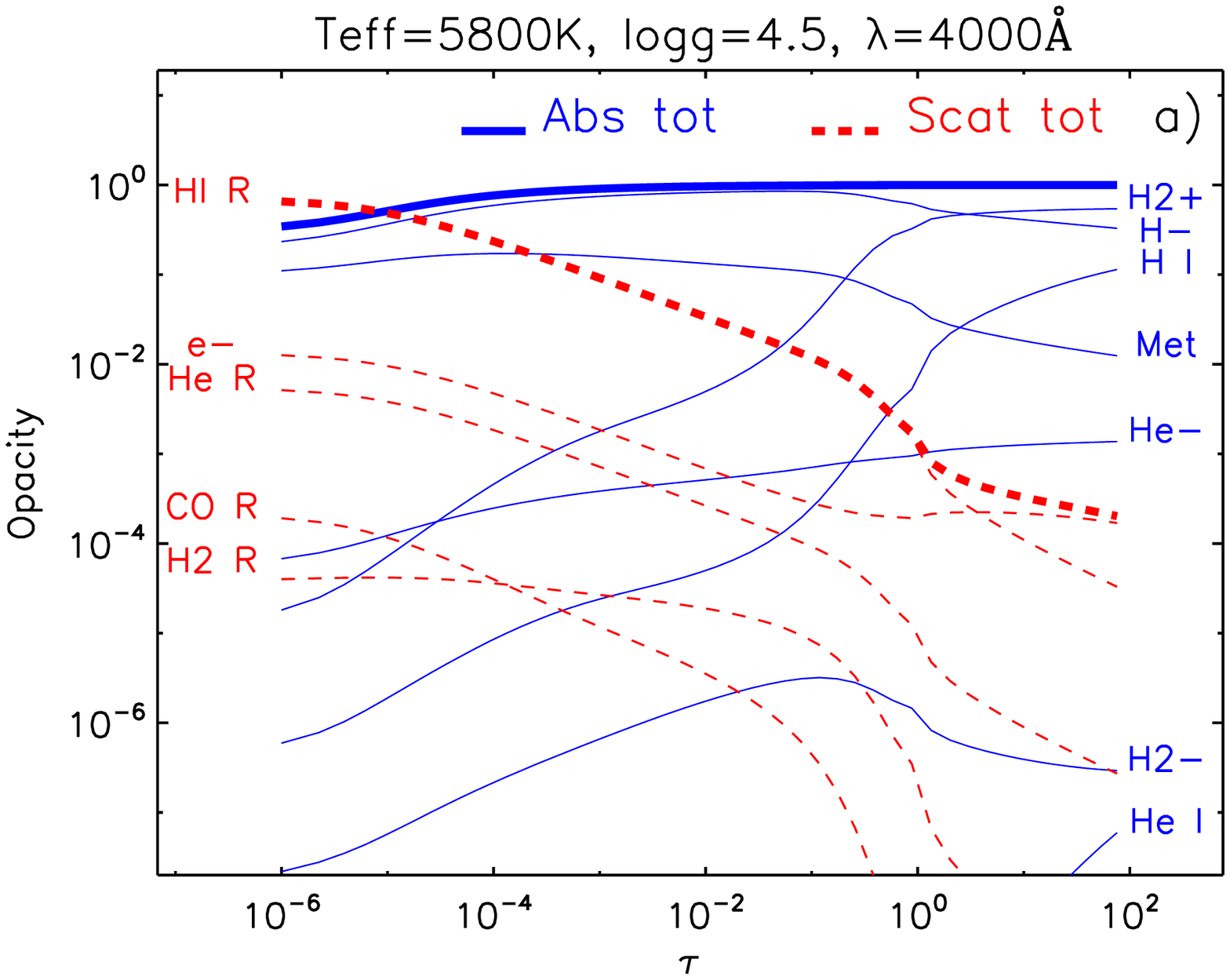}
  \includegraphics[width=.45\linewidth]{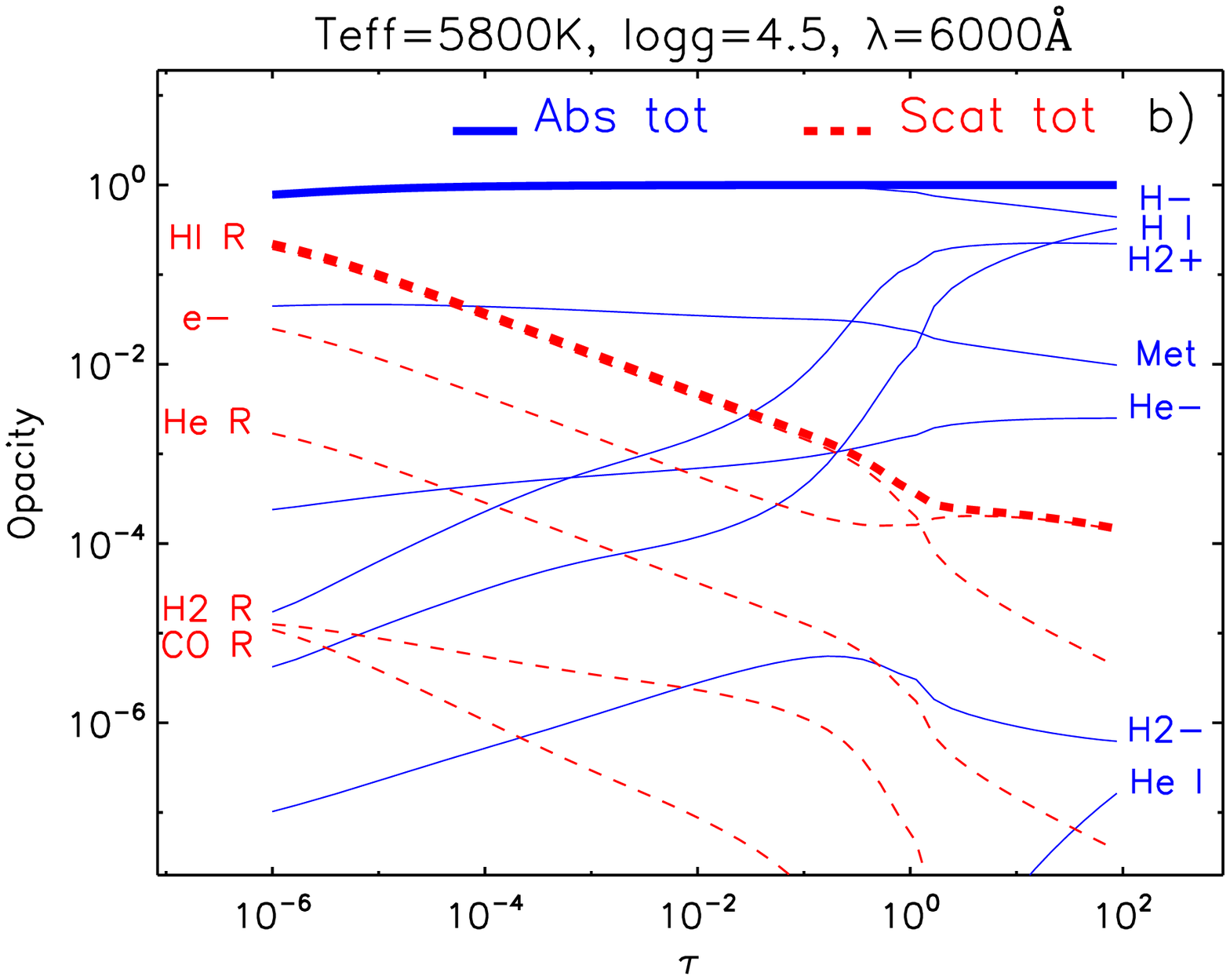}
\end{minipage}%

\begin{minipage}{1.0\textwidth}
  \centering
 \includegraphics[width=.45\linewidth]{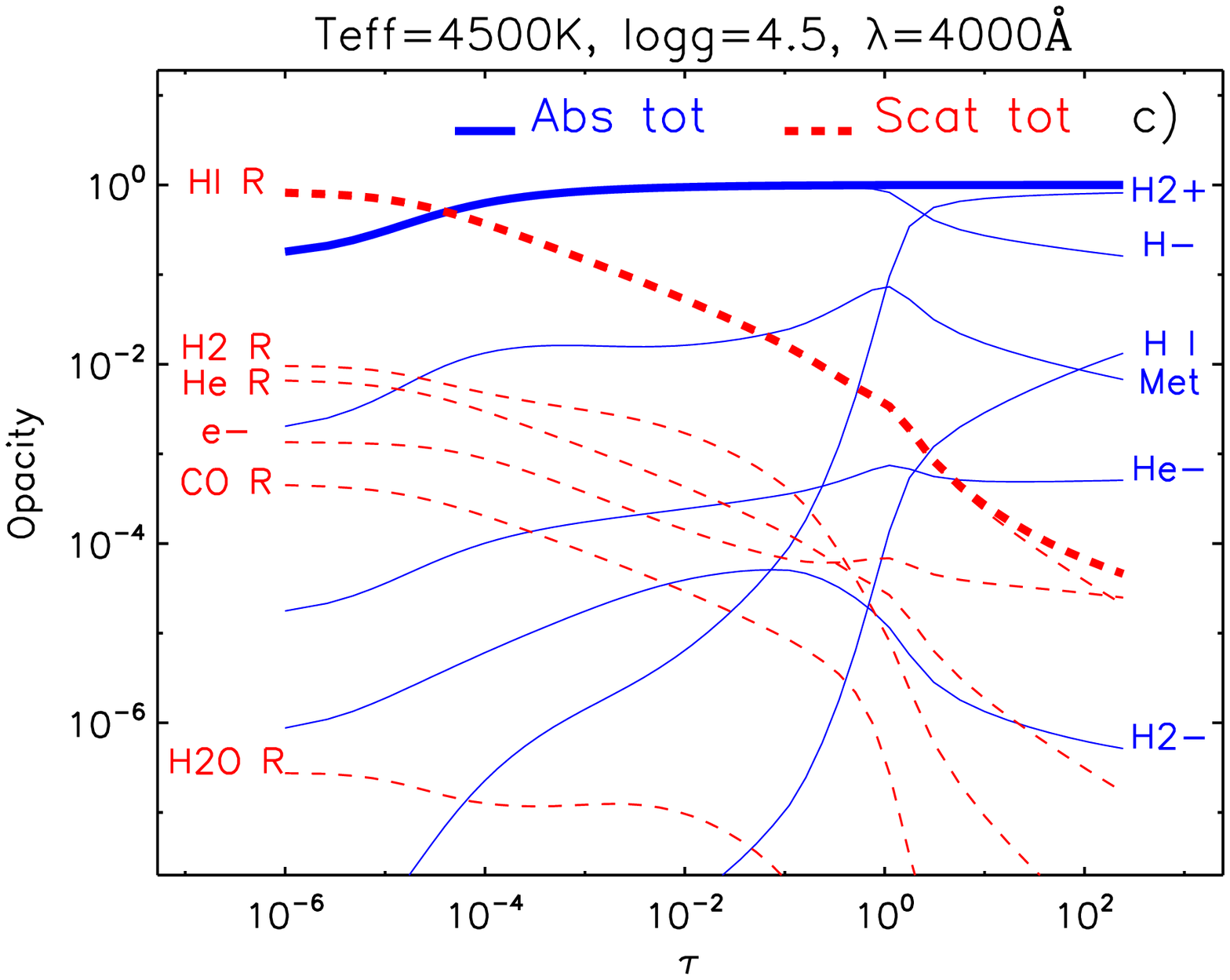}
   \includegraphics[width=.45\linewidth]{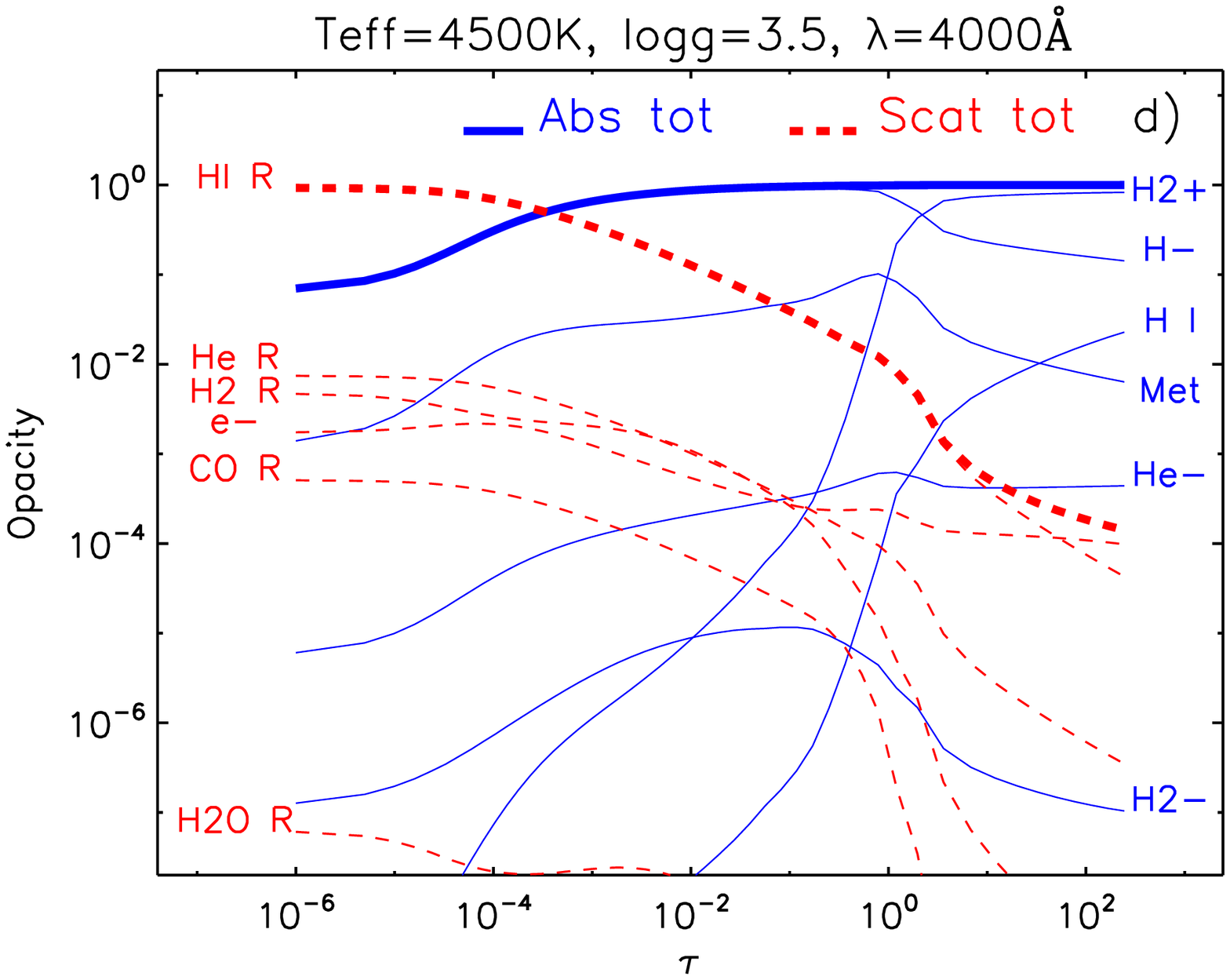}
\end{minipage}  

    \caption{Normalized scattering (dashed lines) and absorption (solid lines) coefficients 
    as a function of optical depth. 
    Titles on each of the panels describe the model parameters for calculation of the opacities.
    Solid lines show all important absorption opacities and dashed lines represent
    all scattering opacities. Thick solid and thick dashed lines are the normalized total absorption and total
    scattering opacities, respectively.}   
   \label{Fig_opac}
   \end{figure*}

Figure \ref{Fig_opac} shows the main contributions to the stellar continuum opacities
(scattering and absorption coefficients normalized to the total opacity) for the different stellar 
atmosphere models with different effective temperatures, gravities, and for wavelengths 
of $4000 \AA$ and $6000 \AA$.  The optical depth scale corresponds to the total opacity at a given wavelength. 

As expected, the most important opacity source 
in the visible wavelengths in the atmospheres of F,G,K stars is absorption by the 
negative ion hydrogen $\rm{H^{-}}$. Another absorption opacity source that plays an important 
role in the deeper layers of stellar atmospheres is $\rm{H_2^+}$, which is a proton-hydrogen atom encounter. 
In addition, in deeper layers of the solar atmosphere $\rm{H I}$ absorption is more noticeable than in cooler
atmospheres, especially at longer wavelengths. 
These opacities act as pure absorption and do not produce any polarization.

The main contribution to the polarization of the stellar continuum
spectrum are Rayleigh scattering on neutral hydrogen $\rm{H I}$ \citep{chandr60}, whish is 
dominated by scattering in the distant line wings of the Lyman series lines \citep{stenflo05} and 
Rayleigh scattering by molecular hydrogen $\rm{H_2}$, which becomes considerable in cooler
stellar atmospheres. Thomson scattering on free electrons is still an
 important source of opacity for the Sun,
 whereas it is completely negligible for cooler stars. 
 
 It is also important to note
  that for the stellar model atmospheres with an
 effective temperature of 4500 K and $\log g = 4.5$ at wavelength of $4000 \AA$ (Fig. \ref{Fig_opac}, c, 
 cooler atmosphere at shorter wavelength) scattering becomes larger than 
 absorption deeper in the atmosphere at $\tau \approx 3\times10^{-4}$, as compared with the solar 
 atmosphere model where it occurs at $\tau \approx 1\times10^{-5}$ (Fig. \ref{Fig_opac}, a). This relative 
 increase of scattering opacity in 
 the stellar atmosphere leads to an absolute increase of the linear polarization in cool atmospheres. 
 
 \subsection {Numerical solution of the radiative transfer problem} 
We have implemented a numerical solution of the radiative transfer problem for polarized light 
in the solar continuum as described in 
\citet{fluri99}. We have developed our code ContPol based on this description and applied it to the stellar 
continuum spectra. Hereafter, we briefly present the formulation of the transfer problem for polarized light. 
For this, we consider a plane-parallel, static atmosphere with homogeneous layers and without 
magnetic fields. The anisotropy that is necessary for scattering polarization is
caused by the center-to-limb variation of the intensity, i.e., temperature gradient within the atmosphere.
 
 Normally, polarized radiation is described by four Stokes parameters $I, Q, U$, and $V$. In our 
 calculations, we choose the coordinate system in such a way that Stokes $Q$ represents 
 the linear polarization in the direction parallel to the stellar limb, and this means 
 that Stokes U is equal to zero. Stokes V is equal to zero because Rayleigh scattering does not 
 produce circular polarization and we neglect magnetic field effects. So the Stokes vector is defined as follows:
 
 \begin{equation}
 \bold{I}_\nu~ = ~ \left( \begin{array}{c}
I \\
Q  \end{array} \right) 
 \end{equation}

The polarized radiative transfer equation in the absence of magnetic fields is written as
 
 \begin{equation}
 \mu~\frac{d\bold{I}_\nu}{d\tau_\nu} = ~\bold{I}_\nu~-~\bold{S}_\nu, 
 \label{rte}
 \end{equation}
 where $\mu = \cos\theta$ defines a line of sight direction with respect to the plane of the atmosphere.
The parameters $\bold{S}_\nu$ and $\bold{I}_\nu$ depend on $\tau$ and $\mu$.
 
 Here, $d\tau_\nu$ is the optical depth defined as
 \begin{equation}
 {d\tau_\nu} = ~-~(k_c~+~\sigma_c)~dz, 
 \end{equation}
 where $k_c$ is the continuum absorption coefficient, $\sigma_c$ is the continuum
  scattering coefficient and $z$ is the geometric height. 
 
 The total source function $\bold{S}_\nu$ is given by
 
 \begin{equation}
 {\bold{S}_\nu} = ~\frac{1}{(k_c~+~\sigma_c)}~(k_c~\bold{B}_\nu~+~\sigma_c~\bold{S}_{s,\nu}), 
 \end{equation}
 where $\bold{B}_\nu$ describes pure absorption and is determined by the Planck function as follows
 
 \begin{equation}
 \bold{B}_\nu~ = ~ \left( \begin{array}{c}
B_\nu(T) \\
0  \end{array} \right), 
 \end{equation}
and  $\bold{S}_{s,\nu}$  expresses the contribution from all radiative sources associated with scattering. It
 can be written as
 
  \begin{equation}
 \bold{S}_{s,\nu}~(\mu)~ = ~\int \bold{P}_R(\mu, \mu ')~\bold{I}_\nu(\mu ')~\frac{d\Omega '}{4\pi}, 
 \end{equation}
 where $\mu '$ is the direction of the incident radiation within the differential solid angle $d\Omega '$. 
 $\bold{P}_R$  is the Rayleigh phase matrix that takes the angular dependence of 
 Rayleigh and Thomson scattering into account and is given by \citet{stenflo94}. 

As in the case of \citet{fluri99}, we first calculate the scattering and 
absorption coefficients, as described in Section 2.1 while neglecting polarization. 
Then the radiative transfer problem for polarized light is solved
with previously computed $k_c$ and $\sigma_c$. To obtain a numerical solution 
of the radiative transfer equations we employ the Feautrier method. This method 
for nonpolarized radiation is discussed by \citet{mihalas78}, while we extend
it to solve the radiative transfer equations for polarized light.

We use the following boundary conditions: the diffusion approximation by \citet{mihalas78}
for Stokes $I$ is assumed at the bottom of the stellar atmosphere, while at the top of the 
atmosphere there is no incoming radiation, i.e., $I = 0$. For both boundary conditions we assume 
no polarization, so the Stokes parameter $Q = 0$. We make several iterations to achieve 
convergence at each level of the atmosphere. Usually more iterations are needed for 
atmospheres with large scattering contributions.

\section{Results}

\subsection{Purely scattering atmosphere}

The classical solution of the radiative transfer equation for an ideal, purely scattering
plane-parallel atmosphere shows the increase of the polarization amplitude up to $11.7\%$
at the very limb of a stellar disk \citep{chandr60}. Pure scattering atmosphere
means that the total opacity is due to scattering and no absorption occurs.
Following \citet{fluri99}, we obtain the pure scattering atmosphere solution by 
artificially redefining the scattering coefficient as the sum of $k_c$ and $\sigma_c$
and setting the absorption coefficient to zero. 
After such assumptions, 
the Stokes $I/I_{center}$ and $Q/I$ components of the outgoing continuum radiation 
field turn out to be independent of frequency and of all thermodynamic properties, so
basically any initial atmosphere can be used for this test.
 
\begin{figure}
   \centering
  \includegraphics[width=\hsize]{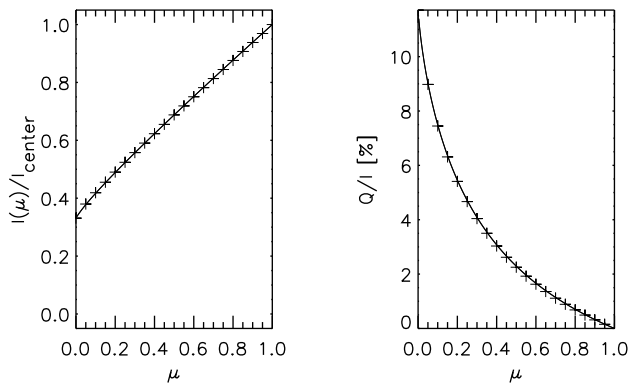}
      \caption{CLVI (left panel)
      and CLVP (right panel) for the 
      exact solution (crosses) given by \citet{chandr60}
      and our calculation of the radiative transfer equation for polarised light for pure scattering 
      atmosphere (solid line). The same solid line is obtained for all wavelengths from $4000 \AA$ 
      to $8000 \AA$ and for different solar and stellar atmospheres assuming pure scattering as described
      in the text.  
              }
         \label{FigChandr}
   \end{figure}

As was shown in \citet{fluri99} all solar model atmospheres give 
identical center-to-limb variations of the polarization and intensity for all 
wavelengths considered, from $4000 \AA$ to $8000 \AA$. With our code we tested 
different solar atmosphere models (FALC, HSRA) and Phoenix atmosphere models
with effective temperatures 5800 K, 4000 K and $\log g = 4.5$
for the $4000 \AA$ to $8000 \AA$ wavelength range. As is shown in Fig.~\ref{FigChandr}, 
we can reproduce precisely Chandrasekhar's solution for pure scattering atmosphere for all considered models. 
This proves that scattering has been correctly calculated in the code. 

\subsection{Limb darkening}

The solar limb darkening was measured many times by different observers who 
fitted the observed center-to-limb variations of the intensity with suitable
analytical functions or limb darkening laws, usually employing up to five 
free parameters that in general depend mostly on wavelength. 

To compare our calculations with the observed solar CLVI,
we have chosen the analytical polynomial function $P_5(\mu)$ given by \citet{neckel94} 
who fitted them to mean continuum measurements, corrected for scattered light.
Any new measurements may differ somewhat from this analytical function because of 
the temporal variability of the limb darkening caused by surface features (e.g., plages). 

\begin{figure*}
   \centering
  \includegraphics[width=\hsize]{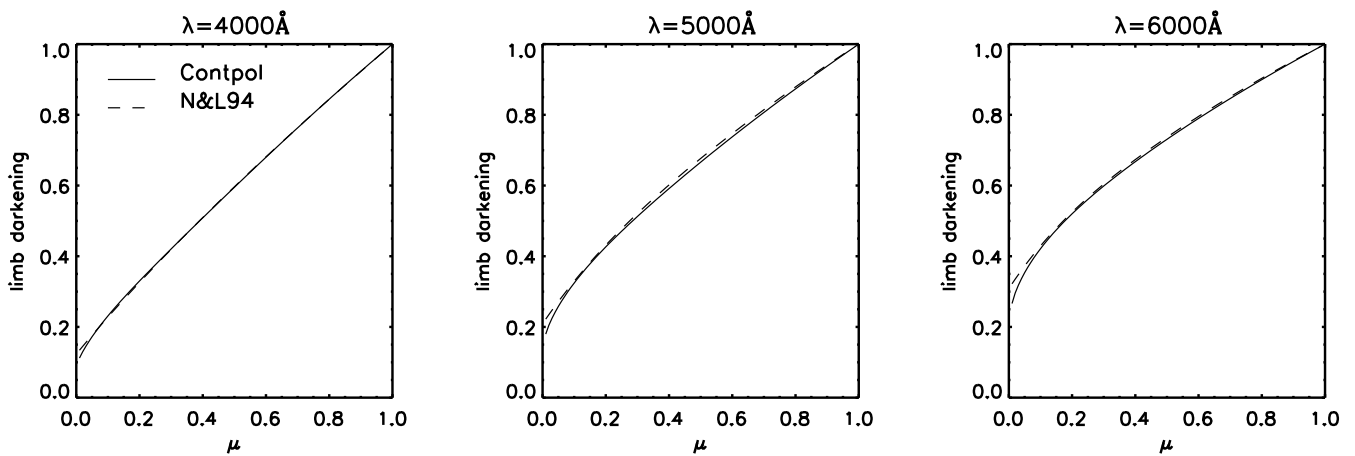}
      \caption{Solar CLVI for different wavelengths. Solid lines 
      in all panels depict our modeled continuum polarization (ContPol) for the HSRA model 
      atmosphere and dashed lines show observations 
      obtained by \citet{neckel94}.}
        \label{FigSolLimbDark}
   \end{figure*}

Figure \ref{FigSolLimbDark} presents observed and calculated solar limb darkening 
for three different wavelengths. As is seen for $\lambda = 4000 \AA$  we have very 
good agreement, while for wavelength of $5000 \AA$ and $6000\AA$ we have small discrepancies. 
Considering natural variations of the solar limb darkening around the \citet{neckel94}
limb darkening polynomial function, we can conclude that limb darkening for the 
Sun is well reproduced by our calculations.  

Likewise, we compared our single wavelength continuum calculations for Phoenix stellar models 
with the results by \citet{claret13} for broadband filters U, B, V. These filters include contributions
from hundreds of angstroms with many spectral lines, especially for cooler atmospheres. 
As expected for hot stars ($>5800K$) 
we can reproduce well broadband simulations since there are not so many lines and 
continuum dominates. With decreasing effective temperatures 
of the stellar models we have larger discrepancies. For cooler atmospheres there are more spectral lines
that contributed to CLVI, therefore the disagreement between our calculations is larger
there. Naturally, to explain broadband observations of the center-to-limb 
variations we need to take all the spectral lines contributing to the bandpass of the filter  into account, 
while our present calculations are useful for explaining the monochromatic measurements at the continuum level.

\begin{figure}[H]
  \centering
   \includegraphics[width=0.95\linewidth]{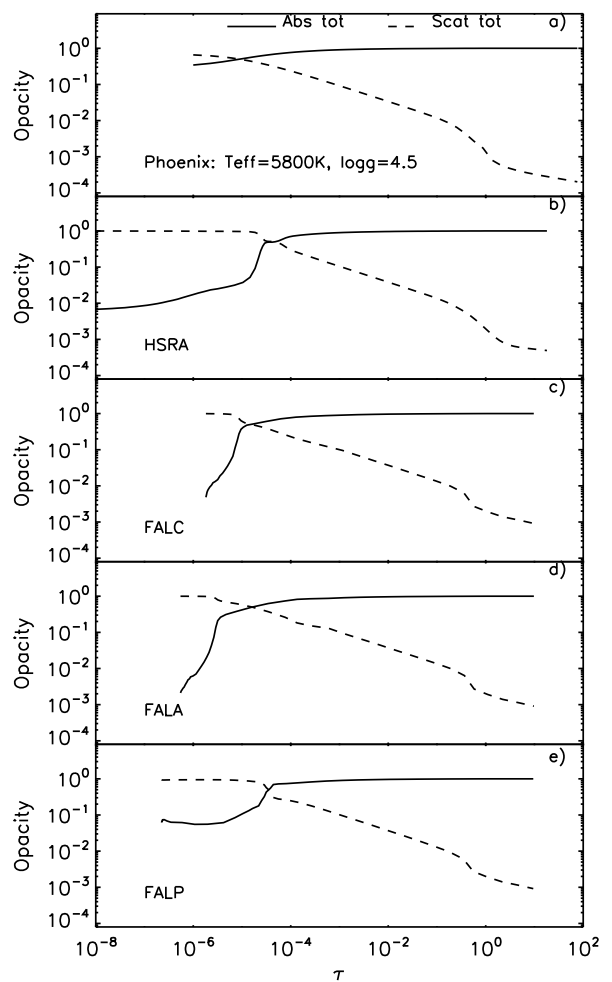}\
  \caption{Normalized scattering (dashed) and absorption (solid) coefficients as a function of optical depth for different 
     solar model atmospheres at $4000 \AA$ wavelength. Different panels correspond to different solar model atmospheres:
     a) Phoenix model with effective temperature of 5800 K and $\log g$ of 4.5; b) HSRA; c) FALC; d) FALA; e) FALP.}   
   \label{Fig_opacsolar}
   \end{figure}

\subsection{Solar limb polarization}

\begin{figure*}
   \centering
  \includegraphics[width=\hsize]{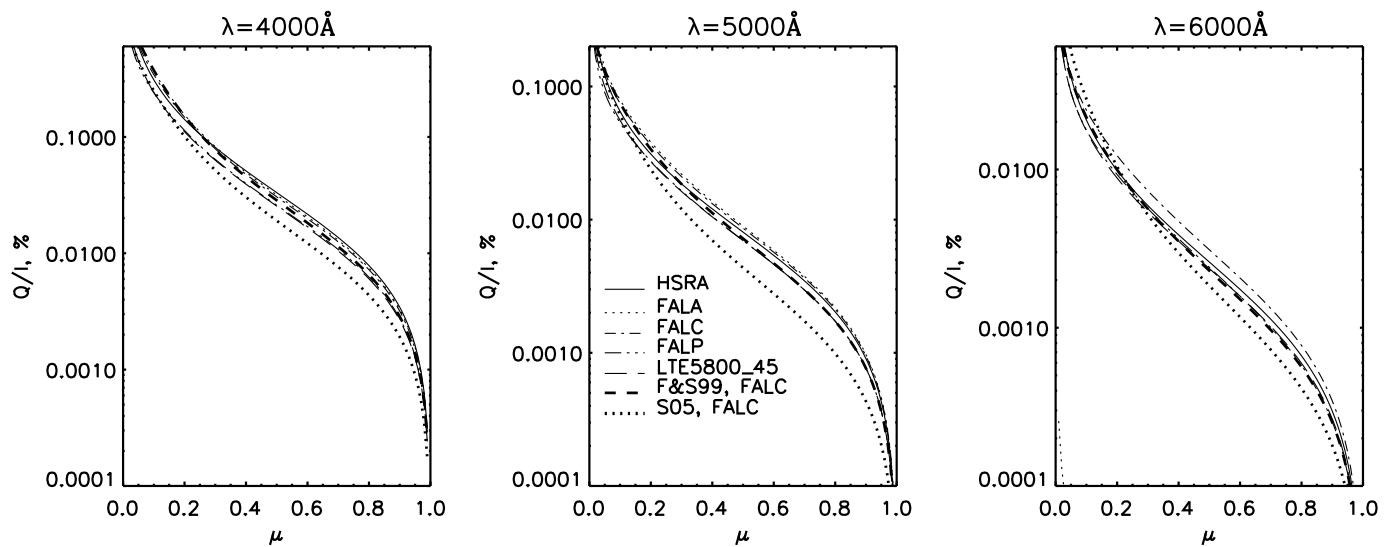}
   \caption{Center-to-limb variation of linear polarization. Different kind of lines 
   depict our calculations for different model atmospheres: FALC, FALP, FALA, HSRA, and 
   LTE5800-45 (Phoenix model for $T_{eff} = 5800K$ and $\log g = 4.5$). The thick dashed line (F\&S99, FALC)
   corresponds to the analytical function for the FALC model taken from \citet{fluri99}, while the
   thick dotted line (S05, FALC) represents the analytical function taken from \citet{stenflo05}.}
      \label{FigSolPol}
   \end{figure*}

We calculate continuum polarization for different solar model atmospheres 
(FALA, FALC, FALP, HSRA, and Phoenix)
and for various wavelengths ($4000 \AA$, $5000 \AA$ and $6000 \AA$).  
We show that the CLVP depends on the model (Fig. \ref{FigSolPol}),
and this is in a good agreement with \citet{fluri99}. 
The difference in polarization at the limb $\mu=0.1$ between FALA (the supergranular cell center) 
and FALP (plage)
models is the largest for all considered wavelengths. For example, for 
wavelength $4000 \AA$ for FALA $Q/I(\mu=0.1) = 0.36\% $, while for FALP
$Q/I(\mu=0.1) = 0.24\% $, which is more than $30\%$ different. Other wavelengths exhibit similar behavior.  
Apart from our calculations, we also show in Fig. \ref{FigSolPol}  
the center-to-limb variation of the continuum linear polarization described by the
analytical approximations taken from \citet[thick dashed line]{fluri99}  
and from \citet[thick dotted line]{stenflo05}. To calculate the continuum polarization with the
empirical equations from both \citet{fluri99} and \citet{stenflo05}, one needs
the solar limb darkening, which they have not given in their papers.
To compare with these two approximations,
we use the limb darkening from our simulation for the FALC model in both cases. To
reproduce the \citet{fluri99} approximation, which depends on the model, 
we choose all parameters for the FALC model from their paper.

To understand the continuum polarization variations for different solar atmosphere models, we analyze 
the opacities for all models presented in Fig. \ref{Fig_opacsolar}. Since the only contributor to linear 
polarization calculations is scattering, the optical depth in the atmosphere 
where scattering becomes dominant determines the amount of observable linear polarization.  
As is seen in Fig.~\ref{Fig_opacsolar}, this depth varies for different solar models. For example, 
the lowest CLVP (Fig.~\ref{FigSolPol}) occurs for the Phoenix
and FALP models. For the Phoenix model the scattering opacity only dominates at the very top of the atmosphere, 
where the number of scattering particles is smaller. However, for both of these models
absorption is still high in the upper levels of the atmosphere, which reduces radiation that can be 
scattered. For the other models the absorption decreases rapidly toward the top of the atmosphere and 
the CLVP is somewhat higher. Opacities variations are due to different temperature 
profiles in these models.

The only way to choose the suitable model is to compare modeled continuum polarization 
with observations.  \citet{stenflo05} derived empirical values of 
continuum polarization on the Sun, which
were extracted from the Second Solar Spectra \citep{gand00, gand02, gand05} with the
 help of a one-parameter model for behavior
of depolarizing lines. He assumed that polarization degrees in the cores of the deepest depolarizing lines  
equal to zero, which, however, may not be fulfilled in blue region. 
 \citet{stenflo05} found that 
his inferred continuum polarization is lower than that predicted by \citet{fluri99}, however, the
difference is within the range of the empirical values. At $4500 \AA$ the lower and upper limit of 
polarization at the limb are $\sim0.07\%$ and $\sim0.12\%$ \citep{stenflo05}, respectively.
\citet{wiehrbiande03} observed the center-to-limb variation of the scattering
polarization in a narrow continuum window at $4506-4508 \AA$ up to the extreme solar limb.
They measured the polarization of $\sim0.12\%$ at disk position $\mu = 0.1$. This is 
 in a very good agreement with \citet{stenflo05}. We believe that the upper limit is more realistic according 
 to the chosen procedure, which is described in \citep{stenflo05} . 
\citet{wiehrbiande03} also
showed that the calculations by \citet{fluri99} for the model FALC fit very well the observations 
up to the limb distance of $\mu = 0.025$, but at exactly $\mu = 0.1$ the measurements have a small discrepancy
from the model and this deviation is within the noise level. 
Modeling scattering polarization in
spectral lines also showed that FALA, FALC, and FALP models lack necessary anisotropy
\citep{bommier06, shapiro11, kleint11}.

\begin{figure*}
   \centering
  \includegraphics[width=\hsize]{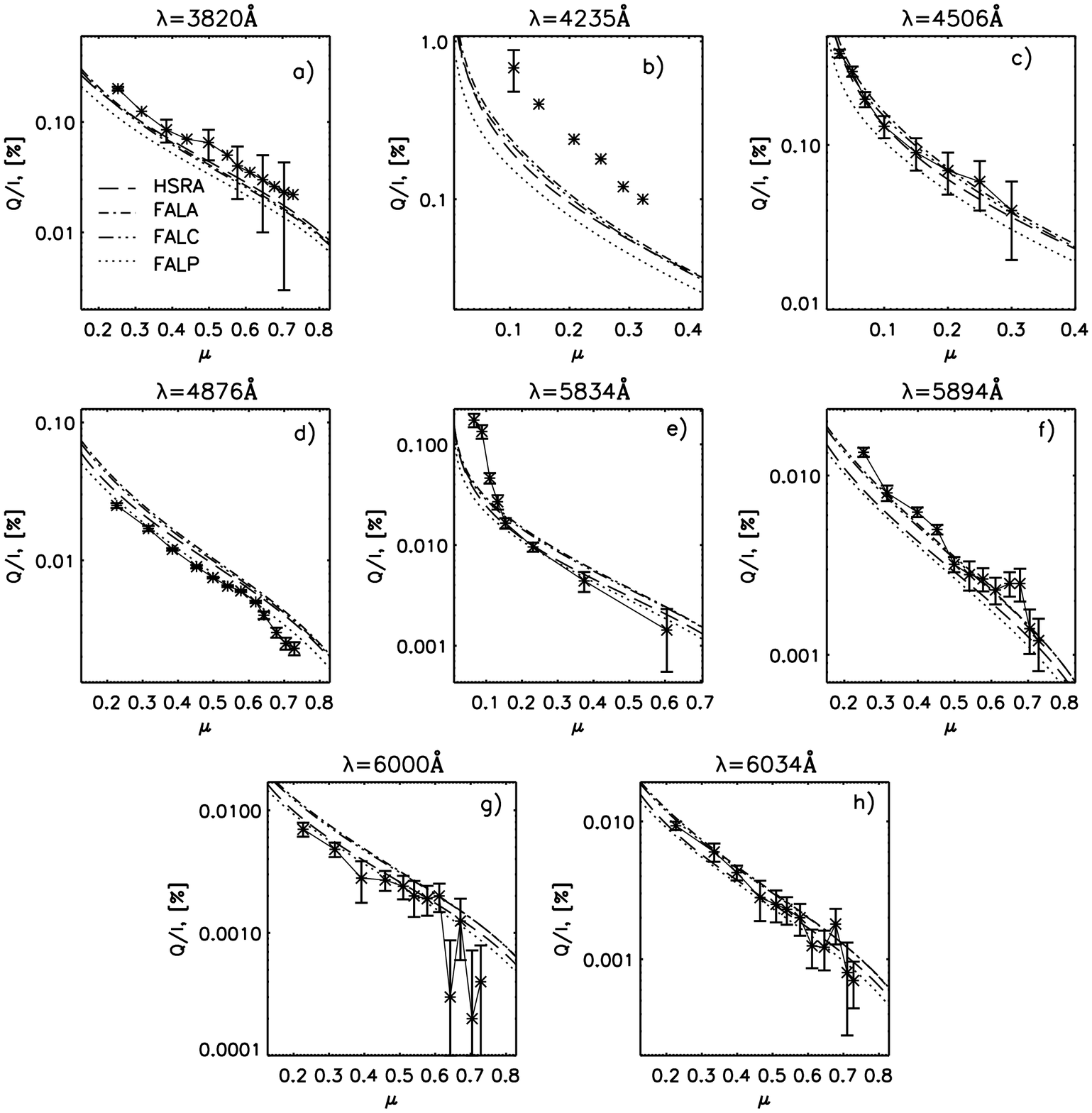}
   \caption{A comparison of modeled and observed center-to-limb variation of linear polarization. 
   Different kind of lines depict our calculations for different model atmospheres: FALC, FALP, FALA, HSRA 
   labeled in plot a). Symbols in a), d), f), g), h) show the solar disk polarization measured by \citet{leroy72}, 
   in c) by \citet{wiehrbiande03}, in b) by \citet{wiehr75}, and in e) by \citet{mickeyorrall74}.
  The wavelength at which the measurement was taken is given in each plot individually.}
      \label{FigPolmeas}
   \end{figure*}

In Figure \ref{FigPolmeas} we compare solar disk polarization measurements from the literature (symbols)
with our modeling for the FALC, FALA, FALP, and HSRA model atmospheres (different line styles). All of the
data were not obtained in the solar continuum, which can lead to a disagreement between observations and calculations caused by line polarization contamination. 
In Figure \ref{FigPolmeas}, a), polarimetric observations
made by \citet{leroy72} in the wing of the Fe I line are presented. They are systematically higher at all limb angles (but are within the error bars for smaller angles) than the calculated continuum polarization for all solar model atmospheres. This is probably because of the Fe I line contribution as well as numerous polarized CN violet system lines in this region \citep{shapiro11}.
A similar situation occurs in Fig. \ref{FigPolmeas}, b), where measurements by \citet{wiehr75} are presented. 
In this case we see the largest discrepancy between
measurements and calculations. However, this discrepany can be explained by a contribution from the nearby strong Ca I line at $\lambda = 4227\rm{\AA}$ blended with Fe I and Fe II lines. The Ca I line has broad wings with a large polarization \citep{gand02}, and the adjacent continuum at 4235\AA\ is apparently contaminated by the line wing . The most recent polarization measurements in the solar continuum ($\lambda = 4506\rm{\AA}$) obtained 
by \citet{wiehrbiande03} are presented in Fig.\ref{FigPolmeas}, c). These data show the best agreement with the solar 
models, except perhaps that shown with the dotted line (FALP model).
The data in Fig. \ref{FigPolmeas}, d) were obtained at 
$\lambda = 4876\AA$. According to the second solar spectrum atlas \citep{gand00}, at this wavelength
there is a slightly depolarizing Fe I line with $Q/I$ of 0.01\% lower than the nearby continuum. Therefore, 
we can expect a somewhat higher polarization in the continuum leading to a better agreement with the solar models.
The next four panels (e, f, g, and h) present the data obtained at longer wavelengths where the polarization is smaller 
and is more difficult to measure. However, despite small disagreements caused by either line contributions or large measurement errors, it appears that our computations for solar models reproduce well the available polarimetric measurements in the continuum made by different authors in different years. It is definitely worth investigating this subject in more detail and obtain more precise measurements at several wavelengths and limb angles, but this is outside the scope of this paper.

\subsection{Stellar limb polarization}

In contrast to numerous investigations of the solar continuum polarization, the limb polarization 
of other stars has not been systematically
studied as much. \citet{harrington70} presented the first calculations of stellar
center-to-limb variation of linear polarization in the case of gray plane-parallel atmosphere.
In this paper we present our calculations of the center-to-limb variation of the
linear polarization in continuum for a large range of stellar models for F, G, K stars in the case of non-gray atmosphere. 

We calculate the radiative transfer equations for the grid of Phoenix model atmospheres within the range of 
effective temperatures from 4500 K to 6900 K at steps of 100K and for $\log g$ from 3.0 to 5.0
at steps of 0.5. For different wavelengths, effective temperatures,
 $\log g$ and various position on the stellar disk $\mu$, we present the value of CLVI 
 in Table 1 and CLVP in Table 2. The Table 1 and 2 for the whole range of the stellar parameters 
 are only available in electronic form at the CDS via anonymous ftp to cdsarc.u-strasbg.fr (130.79.128.5)
 or via http://cdsweb.u-strasbg.fr/cgi-bin/qcat?J/A+A/

In addition, we fit each CLVI with a polynomial and also present 
the table of the polynomial coefficients. The polynomials used are the following:

\begin{equation}
f(\mu) = \sum_{i=0}^{N} a_i *\mu^{i/2}
\end{equation}
where $f(\mu)$ is the CLVI ($I(\mu)/I(1.0)$).
Note that for a good fit to the CLVI, a fourth order polynomial is required. The coefficients of polynomials are only available
in electronic form at the CDS via anonymous ftp to cdsarc.u-strasbg.fr (130.79.128.5)
 or via http://cdsweb.u-strasbg.fr/cgi-bin/qcat?J/A+A/.

  \begin{figure*}
   \centering
  \includegraphics[width=0.98\hsize]{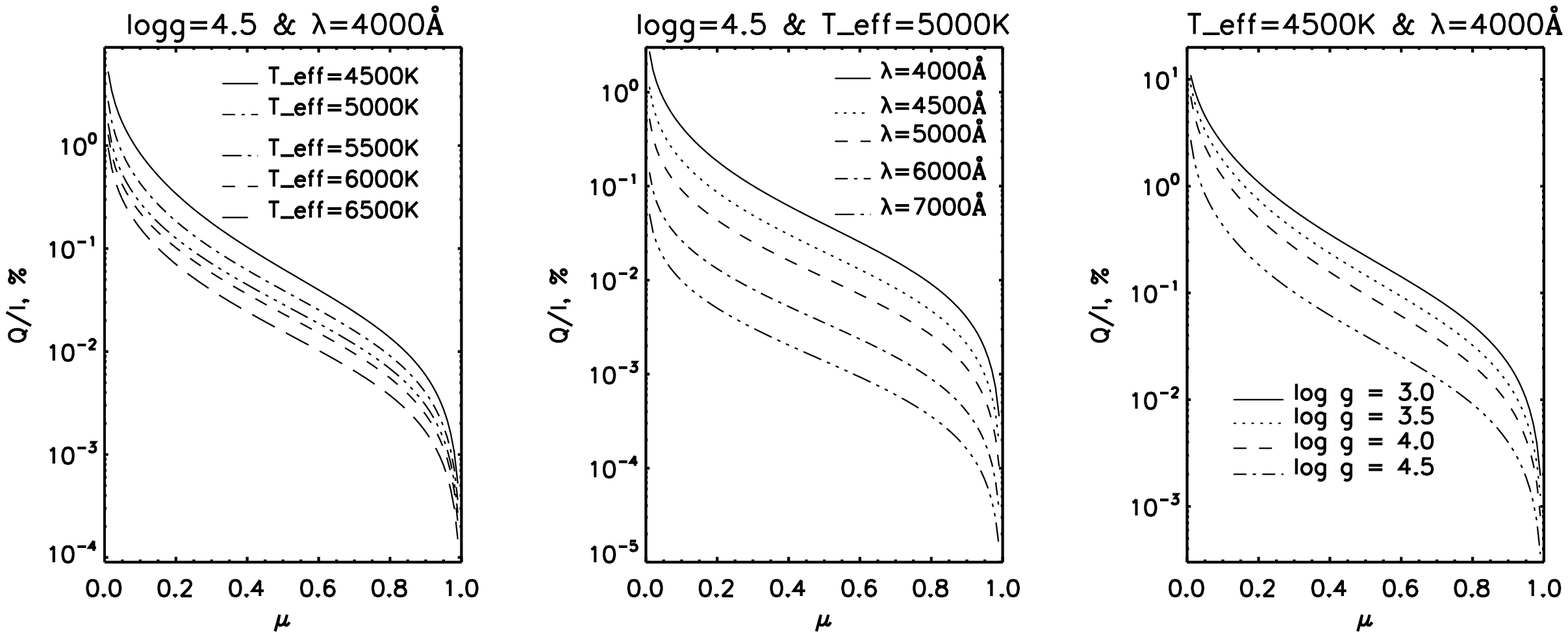}
      \caption{Center-to-limb variation of the continuum polarization for different stellar models. 
      The title above each of the panels indicates the fixed parameters, while labels 
      on each plot describe different curves. Note continuum polarisations are given in logarithmic scale.}
        \label{FigStarPol}
   \end{figure*}

\begin{figure*}
 \centering
 \includegraphics[width=.8\linewidth]{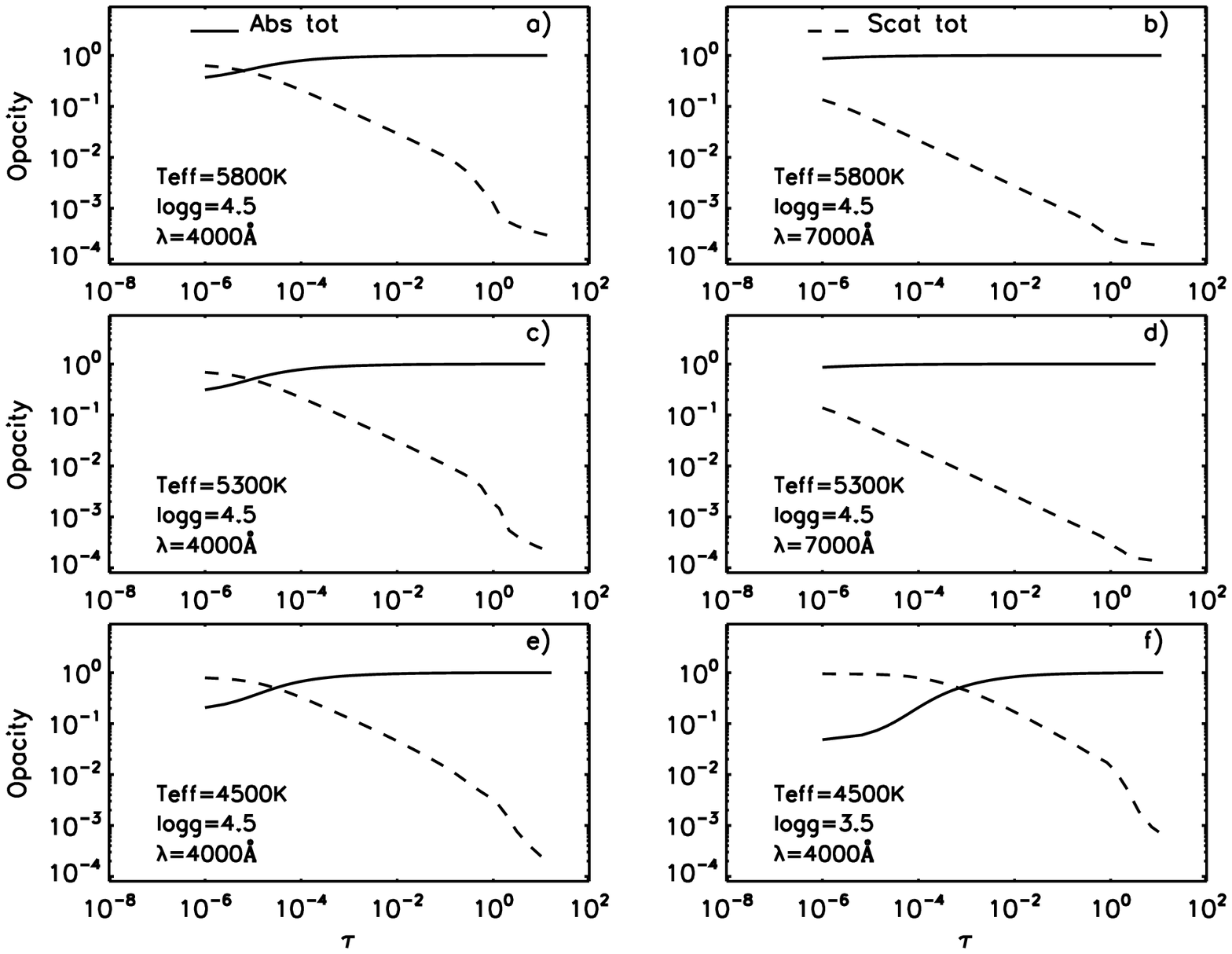}
 
     \caption{Normalized scattering and absorption coefficients as a function of optical depth $\tau$ for Phoenix 
    stellar model atmospheres. Different panels show the opacities for various parameters of the models that
    are described in a legend of each panels. 
    Panels a) and b) present the opacities for stars with effective temperature of 5800K and $\log g$ of 4.5 for
     different wavelengths of $4000\AA$ and $7000\AA$, respectively. Panels c) and d) are for stars
     with the same $\log g$ and at the same wavelengths, respectively but for effective temperature of 5300K.
     Panels e) and f) are for stars with identical effective temperature of 4500K and at the same wavelength of 
     $4000\AA$ except for two different values $\log g$ of 4.5 and 3.0, respectively}
        \label{Fig_opacstellar}
   \end{figure*}

In Fig. \ref{FigStarPol} we show the dependence of stellar continuum polarization on 
the model effective temperature, gravity, and wavelength. As seen in Fig. \ref{FigStarPol} (the first panel)
the cooler the star, the higher the continuum polarization. As the dominant scattering
opacity is due to Rayleigh scattering, higher polarization for shorter wavelengths is obtained (see Fig. \ref{FigStarPol}, second panel). 
The last panel in Fig. \ref{FigStarPol} shows that larger gravity
of a star leads to lower linear polarization. Therefore, the largest continuum 
polarization can be detected on low-gravity cool stars.  
As in the case of the subsection 3.3, we analyze the normalized absorption and
scattering opacities to explain the continuum polarization behavior for different stars.
Figure \ref{Fig_opacstellar} presents opacity calculations for three effective temperatures, two 
$\log g$ values and two different wavelengths. As mentioned above, one of 
the important factors for the continuum polarization 
calculation is the optical depth in the atmosphere where scattering processes dominate over
absorption. Hence, Fig. \ref{Fig_opacstellar} (a, c, e) shows that for the same $\log g$ 
and wavelength region, scattering starts to dominate deeper in the atmosphere for cooler stars than for 
solar-type stars, which leads to increasing linear polarization in cool stars 
(Fig. \ref{FigStarPol}, first panel). We also note 
that for a given wavelength and a given temperature  scattering becomes dominant at a deeper atmospheric 
level for stars with a gravity of $\log g = 3.0$ compared to stars with $\log g = 4.5$ (Fig. \ref{Fig_opacstellar}, e, f).
For the red part of the visual spectrum  $7000 \AA$ (Fig. \ref{Fig_opacstellar}, b), 
where Rayleigh scattering is not so strong, the absorption is the dominant opacity over the entire atmosphere, and
that leads to decreasing linear polarization.

\begin{sidewaystable*}
\caption{\label{CLVI} Calculated center-to-limb variation of intensity for different stellar parameters. All values of $I(\mu)/I(1.0) = 1.0$ at $\mu=1.0$. Complete table is available at the CDS.}
\centering
\begin{tabular}{c c c c c c c c c c c c c}
\hline
\multicolumn{1}{c}{Wavelength, $\AA$} & \multicolumn{1}{c}{T, K} & \multicolumn{1}{c}{$\log g$} & \multicolumn{10}{c}{$I(\mu)/I(1.0)$} \\
\multicolumn{1}{c}{}                   & \multicolumn{1}{c}{}       & \multicolumn{1}{c}{}  	& 0.1     & 0.15     & 0.2 	& 0.25 & 0.3 & 0.35 & 0.4 & 0.5 & 0.7 & 0.9    \\
\hline
4000 & 4500 & 3.0 & 0.1565 & 0.1972 & 0.2412 & 0.2875 & 0.3353 & 0.3838 & 0.4328 & 0.5307 & 0.7233 & 0.9095 \\
4000 & 4500 & 3.5 & 0.1458 & 0.1862 & 0.2296 & 0.2756 & 0.3232 & 0.3719 & 0.4211 & 0.5201 & 0.7160 & 0.9069 \\
4000 & 4500 & 4.0 & 0.1370 & 0.1768 & 0.2196 & 0.2649 & 0.3122 & 0.3609 & 0.4106 & 0.5116 & 0.7131 & 0.9070 \\
4000 & 4500 & 4.5 & 0.1309 & 0.1707 & 0.2136 & 0.2592 & 0.3070 & 0.3564 & 0.4069 & 0.5094 & 0.7132 & 0.9076 \\
4000 & 4500 & 5.0 & 0.1375 & 0.1809 & 0.2266 & 0.2741 & 0.3230 & 0.3729 & 0.4232 & 0.5244 & 0.7228 & 0.9107 \\
4000 & 4600 & 3.0 & 0.1679 & 0.2111 & 0.2570 & 0.3045 & 0.3529 & 0.4017 & 0.4505 & 0.5469 & 0.7339 & 0.9131 \\
4000 & 4600 & 3.5 & 0.1566 & 0.1996 & 0.2451 & 0.2924 & 0.3409 & 0.3900 & 0.4393 & 0.5370 & 0.7274 & 0.9109 \\
4000 & 4600 & 4.0 & 0.1465 & 0.1885 & 0.2333 & 0.2800 & 0.3281 & 0.3771 & 0.4267 & 0.5264 & 0.7231 & 0.9106 \\
4000 & 4600 & 4.5 & 0.1389 & 0.1804 & 0.2245 & 0.2708 & 0.3189 & 0.3682 & 0.4183 & 0.5198 & 0.7203 & 0.9101 \\
4000 & 4600 & 5.0 & 0.1384 & 0.1815 & 0.2272 & 0.2751 & 0.3247 & 0.3754 & 0.4265 & 0.5289 & 0.7275 & 0.9128 \\
\end{tabular}

\vspace{20pt}

\caption{\label{CLVQ} Calculated center-to-limb variation of linear polarization in continuum spectra of different stars. 
All values of $Q/I(\mu) = 0.0$ at $\mu=1.0$. Complete table is available at the CDS.}
\centering
\begin{tabular}{c c c c c c c c c c c c c}
\hline
\multicolumn{1}{c}{Wavelength, $\AA$} & \multicolumn{1}{c}{T, K} & \multicolumn{1}{c}{$\log g$} & \multicolumn{10}{c}{$Q/I(\mu)$} \\
\multicolumn{1}{c}{}                   & \multicolumn{1}{c}{}       & \multicolumn{1}{c}{}  	& 0.1 	 & 0.15     & 0.2 	& 0.25 & 0.3 & 0.35 & 0.4 & 0.5 & 0.7 & 0.9    \\
\hline

4000 & 4500 & 3.0 & 0.0248575 & 0.0158246 & 0.0108887 & 0.0078738 & 0.0058952 & 0.0045257 & 0.0035370 & 0.0022302 & 0.0008833 & 0.0002178\\ 
4000 & 4500 & 3.5 & 0.0175474 & 0.0109756 & 0.0074668 & 0.0053531 & 0.0039794 & 0.0030365 & 0.0023609 & 0.0014765 & 0.0005784 & 0.0001416\\ 
4000 & 4500 & 4.0 & 0.0122426 & 0.0075658 & 0.0051052 & 0.0036373 & 0.0026896 & 0.0020421 & 0.0015800 & 0.0009790 & 0.0003781 & 0.0000919\\ 
4000 & 4500 & 4.5 & 0.0083568 & 0.0050986 & 0.0034093 & 0.0024111 & 0.0017715 & 0.0013376 & 0.0010300 & 0.0006334 & 0.0002425 & 0.0000587\\ 
4000 & 4500 & 5.0 & 0.0049754 & 0.0029966 & 0.0019937 & 0.0014091 & 0.0010370 & 0.0007851 & 0.0006065 & 0.0003753 & 0.0001451 & 0.0000353\\ 
4000 & 4600 & 3.0 & 0.0214401 & 0.0136749 & 0.0094553 & 0.0068754 & 0.0051755 & 0.0039930 & 0.0031349 & 0.0019924 & 0.0007987 & 0.0001986\\ 
4000 & 4600 & 3.5 & 0.0150665 & 0.0094382 & 0.0064477 & 0.0046461 & 0.0034716 & 0.0026620 & 0.0020793 & 0.0013110 & 0.0005202 & 0.0001285\\ 
4000 & 4600 & 4.0 & 0.0105818 & 0.0065487 & 0.0044335 & 0.0031728 & 0.0023579 & 0.0017993 & 0.0013989 & 0.0008742 & 0.0003420 & 0.0000839\\ 
4000 & 4600 & 4.5 & 0.0072985 & 0.0044702 & 0.0030042 & 0.0021367 & 0.0015789 & 0.0011985 & 0.0009273 & 0.0005747 & 0.0002223 & 0.0000542\\ 
4000 & 4600 & 5.0 & 0.0047062 & 0.0028453 & 0.0018947 & 0.0013379 & 0.0009832 & 0.0007432 & 0.0005733 & 0.0003542 & 0.0001369 & 0.0000334\\ 

\end{tabular}
\end{sidewaystable*}


\section{Summary}

     We have solved the radiative transfer equations for polarized light 
     in the continuum spectra accounting for absorption, Rayleigh, and Thomson scattering. 
     Thomson scattering is more important for hotter stellar atmospheres, while for almost all stellar
     atmospheres considered here the most important opacities are Rayleigh scattering on neutral hydrogen 
     and absorption by a negative hydrogen ion. For cooler atmospheres, Rayleigh scattering on molecular hydrogen becomes 
     more prominent but is still not dominating in the considered range of the effective temperatures of a star.

     We present the results of our calculations of the CLVI and CLVP in the stellar continuum
     spectra with effective temperatures 4500 K - 6900 K and gravities $\log g = 3.0 - 5.0$ 
     and in the spectral range from $4000 \AA$ to $7000 \AA$.
     Since Rayleigh scattering is dominant, the scattering polarization is larger in the blue wavelengths.

     The deeper in the atmosphere the scattering opacity becomes dominant over the absorption opacity, the 
      larger the linear polarization can be observed. Since in low-gravity cool stars scattering becomes dominant  
      very deep in the atmosphere, the CLVP is larger for cool giants. So,  
      low gravity cool stars have much larger polarization in their continuum spectra
      although they have smaller CLV of the intensity. 
     
      We conclude that linear polarization is a sensitive tool 
      to test stellar model atmospheres, with respect to their temperature profiles and particle density distribution.
      In addition, for testing the stellar models, the  CLVP is very useful for calculations of stellar disk symmetry breaking effects, 
      such as planetary transits and presence of spots on the stellar disk, which we will present in a forthcoming paper.

\begin{acknowledgements}
      This work was supported by the HotMol project funded by the ERC Advanced Grant.
      We thank  Martin K\"{u}rster for the comments and suggestions and an anonymous 
      referee for comments that improved this paper.

\end{acknowledgements}

\bibliographystyle{aa} 
\bibliography{Cont.bib} 
  
\end{document}